\documentclass[useAMS]{mn2e}
\usepackage{amsmath}
\usepackage{textcomp}
\usepackage{amssymb}
\usepackage[pdftex]{graphicx}
\usepackage{amssymb}

\usepackage[margin=.8in, paperwidth=8.5in, paperheight=11in]{geometry}
\usepackage{multicol}
\usepackage{epsfig}
\usepackage{graphicx}
\usepackage{dblfloatfix}

\begin{document}

\title{Advection of Matter and B-Fields in Alpha-Discs}
\author[S. Dyda et al.]
{\parbox{\textwidth}{S.~Dyda,$^{1}$
R.V.E.~Lovelace,$^{2}$
G.V.~Ustyugova,$^{3}$
P.S. Lii,$^{2}$,
M.M.~Romanova,$^{2}$\\
 \& A.V.~Koldoba$^{3}$}\vspace{0.3cm}\\
 $^{1}$Department of Physics, Cornell University, Ithaca, NY 14853:email: sd449@cornell.edu\\
 $^{2}$Department of Astronomy, Cornell University, Ithaca, NY 14853\\
$^{3}$Keldysh Institute for Applied Mathematics, Moscow, Russia\\
}

\date{\today}
\pagerange{\pageref{firstpage}--\pageref{lastpage}}
\pubyear{2012}

\label{firstpage}

\maketitle

\begin{abstract}

   We have carried out and analyzed  a set of axisymmetric MHD simulations
of the evolution of a turbulent/diffusive accretion disc around an
initially unmagnetized star.  The disc is
initially threaded by a weak  magnetic field where 
the magnetic pressure is significantly less than the kinetic
pressure in the disc.
The viscosity and magnetic diffusivity are modeled by
two  ``alpha''  parameters, while the coronal region above 
the  disc is treated using ideal MHD.   
      The  initial  magnetic field is taken to consist of three poloidal field loops threading the disc.
The  motivation for this study is to understand the advection of disc 
matter and magnetic field by the turbulent/diffusive disc.  
    At early times ($ \lesssim 400$ orbits of the inner disc), the innermost
field loop twists and its field lines become open.  The twisting of
the opened field lines leads to the formation of both an inner collimated, magnetically-dominated jet, and at larger distances from the axis a matter dominated uncollimated wind.   For later times ($>1000$), the strength of the magnetic field decreases owing to field reconnection and
annihilation in the disc.
   For the early times, we have derived from the simulations both the matter accretion speed in the disc $u_{\rm m}$ and the accretion speed of the magnetic field $u_B$ which is determined by measuring the speed 
 of the inward motion of the inner O-point of the magnetic field in
 the equatorial plane.  
     We show that the derived  $u_{\rm m}$  agrees approximately with the predictions of a model where the accretion speed is the sum of two 
 terms, one due to  the disc's viscosity (which gives a radial outflow of angular momentum in the disc), and a second  due to the twisted magnetic field at the disc's surface (which gives a vertical outflow of angular momentum).
   At later times the magnetic contribution to $u_{\rm m}$ becomes
small compared to the viscous contribution.
   For early times we find that $u_{\rm m}$ is larger than the magnetic
field accretion speed $u_B$ by a factor of $\sim 2$ for the case where
the alpha parameters are both equal to $0.1$.

\end{abstract}

\begin{keywords} accretion,  accretion discs -- MHD -- black hole physics, magnetic fields, jets, stars: winds, outflows
\end{keywords}

\section{Introduction}

         Early studies of the advection and diffusion of a
large-scale magnetic field threading a turbulent
disc indicated that a {\it weak} large-scale field would
diffuse outward rapidly (van Ballegooijen 1989; 
Lubow, Papaloizou, \& Pringle 1994;
Lovelace, Romanova, \& Newman 1994; Lovelace,
Newman, \& Romanova 1997).
       This rapid outward diffusion may however be offset by
the highly conducting surface layers of the disc where
the magnetorotational instability (MRI) and associated 
turbulence is suppressed (Bisnovatyi-Kogan \& Lovelace 2007; 
Rothstein \& Lovelace 2008).  The magnetic field is 
``frozen-in'' in the conducting surface layers which tend to flow 
inward at approximately the disc accretion speed.
     This conclusion is supported by an analytic model for the vertical
 profiles of the velocity and field components 
of a stationary accretion disc  developed by Lovelace, Rothstein, and Bisnovatyi-Kogan (2009).   
     This model predicts that the
inward or outward transport  of the poloidal magnetic flux  is determined
by both the plasma $\beta_0$ (the ratio of the midplane plasma pressure
to the midplane magnetic pressure) and the efficiency of the magnetic
disc wind in removing angular momentum from the disc
(Bisnovatyi-Kogan \& Lovelace 2012).
Guilet and  Ogilvie (2012, 2013)
independently developed an analytic model for the vertical structure
of a turbulent/diffusive disc threaded by a large scale magnetic field,
and they find a reduction in the rapid outward field diffusion.

   Accretion discs around black holes are considered in many cases to be
threaded by a large-scale  magnetic field (Lovelace 1976). 
      This field may be transported inward  from the interstellar medium
by the accreting disc plasma (as investigated here), or it may arise from
dynamo activity in the disc (e.g., Pariev,  Colgate, \& Finn 2007).
   Of course the discs around magnetized stars may be threaded at large distances by the disconnected stellar magnetic field (Lovelace, Romanova, \& Bisnovatyi-Kogan 1995).
     The large-scale field may be in the form of  magnetic loops threading the disc and extending into a low density plasma corona as sketched in Figure 1.
Differential rotation of the disc acts to open the magnetic
loops which have footpoints at different radii (Newman, Newman, \& Lovelace 1992).  Also, the differential rotation acts to
give an axisymmetric field.
Such large scale magnetic fields can have an essential role in forming
jets and winds.

     In  previous axisymmetric magnetohydrodynamic (MHD) simulations of a disc threaded by magnetic loops,  the disc was
 treated   as  a conducting boundary condition with plasma outflow and Keplerian azimuthal velocity (Romanova et al. 1998).   
    These simulations showed  that  the innermost loop  inflates
and opens significantly faster than the outer loops due to the larger differential rotation of the disc close to the star.  The opened magnetic fields carry away energy, angular momentum, and mass from the disc.
     One or more neutral layers  form between the regions of oppositely directed
magnetic  field lines leading to  field reconnection and annihilation.

    The aim of the present work is to understand the  dynamics of the
magnetic field loops {\it and} the dynamics of the disc in response
to the twisting of the loops.  
The magnetic field mediates
an outflow of energy, angular momentum and matter from the disc to
a jet and wind.  
     At the same time the disc accretion rate can
be strongly  enhanced by the
angular momentum outflow to the jet or wind.
    We treat the disc as
a viscous/diffusive plasma taking  into account fully  
the back reactions of the coronal field  on the disc.   
    The turbulent viscosity $\nu_t$  of the disc is modelled  with  an $\alpha_\nu$ coefficient using the  Shakura and Sunyaev (1973)
prescription.   
     The turbulent magnetic diffusivity $\eta_t$ is modeled with a second $\alpha_\eta$ coefficient as proposed by
Bisnovatyi-Kogan and Ruzmaikin (1976).
    The viscosity and diffusivity  are assumed to arise from 
turbulence triggered by the magneto-rotational instability inside
the disc (Balbus \& Hawley 1998), but this
turbulence is not modeled in the present simulations. 
   The low density coronal plasma outside the disc is treated using ideal MHD.

    We carry out axisymmetric simulations using a
Godunov-type scheme to solve the MHD equations, including  
viscosity and magnetic diffusivity inside the disc as described by
Ustyugova et al. (2006). 
    Our initial magnetic field configuration consists of three loops in the simulation region.   New unmagnetized matter is supplied to
the disc at the outer boundary. 
    
For this configuration the innermost loop opens up rapidly and
forms a collimated magnetically dominated jet near the $z-$axis and
an uncollimated matter dominated wind at larger distances from
the axis.
    The second loop opens at a later time due to the smaller shear
in the disc and it produces a matter
 dominated wind.  
       The outermost loop
reconnects before there is time for it to open.

     The loop configuration allows us to evaluate {\it both} the accretion
speed of the magnetic field and accretion speed of the disc matter.
This allows a comparison with the analytic model of field accretion
of Lovelace et al. (2009).

  The paper is organized as follows:  We discuss the setup for our simulations including the  initial 
and boundary conditions in Sec. 2.
     Section 3 describes our results on the dynamics of the disc, the generation of  jets and disc winds, and the accretion speeds of
the matter and magnetic field.   Section 4 gives the
conclusions of this work.

\begin{figure}
                \centering
                \includegraphics[width=.4\textwidth]{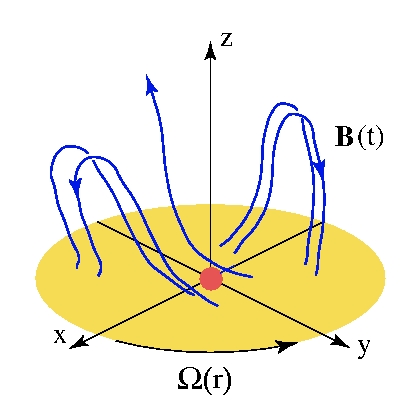}
        \caption{Sketch of an accretion disc threaded by open and closed magnetic field lines.
        Differential rotation of the disc will act to rapidly  give an axisymmetric field
configuration considered in this work.}
\label{fig:cartoon}
\end{figure}

\section{Theory}

\subsection{Basic Equations}

    The plasma flows are assumed to be described by the
equations of non-relativistic magnetohydrodynamics (MHD). 
In a non-rotating reference frame the equations are
\begin{subequations}
 \begin{equation}
\frac{\partial \rho}{\partial t} + \nabla \cdot \left( \rho \mathbf{v} \right) = 0~, 
\end{equation}
\begin{equation}
 \frac{\partial \rho \mathbf{v}}{\partial t} + \nabla \cdot \mathcal{T} = \rho \mathbf{g}~,
\end{equation}
\begin{equation}
\frac{\partial \mathbf{B}}{\partial t} + c\: \nabla \times \mathbf{E} = 0~,
\end{equation}
\begin{equation}
\frac{\partial \left( \rho S \right)}{\partial t} + \nabla \cdot \left( \rho  \mathbf{v}S \right) = \mathcal{Q}~.
\end{equation}
\end{subequations}
Here, $\rho$ is the mass density, $S$ is the specific entropy, $\mathbf{v}$ is the flow velocity, $\mathbf{B}$ is the magnetic field, $\mathcal{T}$ is the momentum 
flux density tensor, $\mathbf{E}$ is the electric field, $\mathcal{Q}$ is the rate of change of entropy per unit volume due to viscous
and Ohmic heating in the disc, and $c$ is the speed of light. 
    We assume that the heating is offset by radiative cooling so that
$Q=0$.
   Also, $\mathbf{g} = -\left[GM/(r-r_S)^2 \right]\hat{r}$ (with
$r_S\equiv 2GM/c^2$)  is the gravitational acceleration due to the
central mass $M$ with the Paczy\'nski-Wiita (1980) correction relevant to neutron stars.
We model the plasma as a non-relativistic ideal gas with equation of state
\begin{equation}
 S = \ln\left( \frac{p}{\rho^{\gamma}}\right)~,
\end{equation}
where $p$ is the pressure and $\gamma = 5/3$.

In most of this paper we use  spherical $(r,\theta,\phi)$ coordinates. 
However,  for some purposes cylindrical coordinates are advantageous,
and they are denoted $(R,\phi,Z)$.

   Both the viscosity and the magnetic diffusivity of the
disc plasma are thought to be due to turbulent fluctuations of the velocity and magnetic field.   
      Outside of the disc,  the
plasma is considered ideal with negligible viscosity and diffusivity.
      The turbulent coefficients are parameterized using
the  $\alpha$-model of Shakura and Sunyaev (1973).
    The turbulent kinematic viscosity is
\begin{equation}
 \nu_t = \alpha_{\nu} \frac{c_s^2}{\Omega_K}~,
\end{equation}
where $c_s$ is the midplane  sound speed, $\Omega_K$ is the Keplerian angular velocity at the given radii and $\alpha_{\nu}\leq 1$ is a dimensionless constant.
 Similarly,  the turbulent magnetic diffusivity is
\begin{equation}
 \eta_t = \alpha_\eta \frac{c_s^2}{\Omega_K}~,
\end{equation}
where $\alpha_\eta$ is another dimensionless constant. 
   The ratio, 
\begin{equation}
 \mathcal{P} = \frac{\alpha_{\nu}}{\alpha_{\eta}}~ ,
\end{equation}
is the magnetic Prandtl number of the turbulence in the disc
which is expected to be of order unity (Bisnovatyi-Kogan
\& Ruzmaikin 1976). 
   Shearing box simulations of MRI driven MHD turbulence
in discs indicate that ${\cal P}\sim 1$ (Guan \& Gammie 2009).

The momentum flux density tensor is given by
\begin{equation}
 \mathcal{T}_{ik} =p\delta_{ik}+ \rho v_i v_k  + \left( \frac{\mathbf{B}^2}{8\pi}\delta_{ik} - \frac{B_i B_k}{4\pi} \right) + \tau_{ik}~,
\end{equation}
where $\tau_{ik}$ is the viscous stress contribution from the turbulent fluctuations of the velocity and magnetic field. 
    As mentioned we assume that these can be represented in the same
way as the collisional viscosity by substitution of the turbulent viscosity.       Moreover, we assume that the viscous stress is determined mainly by the
gradient of the angular velocity because the azimuthal velocity is the dominant velocity of the disc. 
   The leading order contribution to the momentum flux density from 
turbulence is therefore
\begin{subequations}
 \begin{equation}
  \tau_{r\phi} = -\nu_t \rho r \sin \theta \frac{\partial \omega}{\partial r}~,
 \end{equation}
\begin{equation}
 \tau_{\theta \phi} = - \nu_t \rho \sin \theta \frac{\partial \omega}{\partial \theta}~,
\end{equation}
\end{subequations}
where $\omega = v_{\phi}/r \sin \theta$ is the plasma angular velocity.

    The transition from the viscous-diffusive disc to the ideal plasma
corona is handled  by multiplying the viscosity and
diffusivity by a dimensionless factor $\xi(\rho)$ which varies from
$\xi =0$ for $\rho\leq 0.25 \rho_d$ to $\xi =(4/3)(\rho/\rho_d -0.25)$
for $0.25\rho_d  < \rho < \rho_d$ to $\xi =1 $ for $\rho >\rho_d$
 (see Appendix B of Lii, Romanova, \& Lovelace 2012).  
   The disc half-thickness $h$ is taken to be the vertical distance
from $Z=0$ to the $0.5\rho_d$ surface.

\subsection{Initial Conditions}

\subsubsection{Initial Magnetic Field}

   The initial magnetic field is described in cylindrical coordinates. 
This field is taken to be force-free in the sense that ${\bf J \times B}=0$ 
in the region $|Z| >0$ with
${\bf B}=(B_R, B_\phi,B_Z)$, where $B_R = -R^{-1}\partial \Psi/\partial Z$
and $B_Z = R^{-1}\partial \Psi/\partial R$.
     Here, $\Psi(R,Z)$ is the flux
function which labels the field lines, ${\bf B\cdot \nabla}\Psi=0$.
This  function satisfies  the Grad-Shafranov 
equation,
\begin{equation}
 \Delta^*\Psi(R,Z) = -H(\Psi){dH(\Psi)\over d\Psi}~,
\label{eq:GS}
\end{equation}
where 
\[\Delta^{*} = \frac{\partial^2}{\partial R^2} - \frac{1}{R}\frac{\partial}{\partial R} + \frac{\partial^2}{\partial Z^2}~,\]
\noindent and $H=H(\Psi)=RB_\phi(R,Z)$ is the poloidal current function (Lovelace et al. 1986).
   For simplicity we take the poloidal current to be proportional
to the flux, 
$ H(\Psi) = k \Psi,$
where $k$ is a constant (Newman et al. 1992).    
The relevant solution to equation (8) is 
\begin{equation}
\Psi(R,Z) = A_1 R \: J_1(a_1R)e^{-b_1|Z|} + A_2 R \: J_1(a_2R)e^{-b_2|Z|}~,
\end{equation}
where $J_1$ is a Bessel function of the first kind, $A_1$, $A_2$ are integration constants and 
$ a_i = (k^2+b_i^2)^{1/2}$ for  $ i=1,2$.
 Qualitatively, the field appears as a number of loops threading the accretion disc. We have chosen the parameters $a_i$ such that three loops fit in our simulation region.
 
The solution (9) is valid in the region $|Z| > 0$ and assumes that 
initially there is a thin current carrying disc in the $Z=0$ plane.
    The cusp in the initial field at $Z=0$ disappears rapidly 
in a time $t_{\rm init} \sim h^2/\eta_t = (\alpha_\eta \Omega_K)^{-1}=
P_0(2\pi\alpha_\eta)^{-1}(r/r_0)^{3/2}$
due to the diffusivity of the disc.        Here, $r_0$ is the reference
radius and $P_0$ the period of the Keplerian orbit at this radius as
discussed in \S 2.4. 
We have assumed $h/R \approx c_s/v_K$ which neglects the
magnetic compression of the disc discussed by Wang, Sulkanen,
and Lovelace (1990).
    For most of the range of $r/r_0$ this time is much smaller than
the field evolution time scale.

\subsubsection{Matter Distribution}
Initially the matter of the disc and corona are assumed to be in mechanical equilibrium (Romanova et al. 2002). 
The initial density distribution is taken to be barotropic with
\begin{equation}
  \rho(p) =
  \begin{cases}
   p/T_{\rm{disc}} & p>p_b ~~{ \rm and} ~~ r \sin \theta \geq r_b~, \\
   p/T_{\rm{cor}} & p<p_b ~~ {\rm or} ~~ r\sin \theta \leq r_b~,
  \end{cases}
\end{equation}
where $p_b$ is the level surface of pressure that separates the cold matter of the disc from the hot matter of the corona and $r_b$ is the initial
value of the  inner radius of the disc.   At this surface
the density has an initial step discontinuity from value $p/T_{\rm{disc}}$ to $p/T_{\rm{cor}}$.

Because the density distribution is barotropic, the initial angular velocity is a constant on coaxial cylindrical surfaces about the $z-$axis. Consequently, the pressure 
can  be determined from  Bernoulli's  equation,
\begin{equation}
 F(p) + \Phi + \Phi_c = \rm{const}~,
\end{equation}
where $\Phi = -GM/|r-r_c|$ is the gravitational potential 
with the Paczy\'nski-Wiita correction, $\Phi_c = \int_{r\sin \theta}^{\infty}\xi d\xi ~\omega^2(\xi)$ is the 
centrifugal potential, which depends only on cylindrical radius $R = r\sin \theta$, and 
\begin{equation}
  F(p) =
  \begin{cases}
   T_{\rm{disc}}\ln(p/p_b) & p>p_b ~~ {\rm and} ~~ r\sin \theta \geq r_b~, \\
   T_{\rm{cor}}\ln(p/p_b) & p<p_b ~~{\rm or} ~~ r\sin \theta \leq r_b~.
  \end{cases}
\end{equation}

\subsubsection{Angular Velocity}

Initially the inner edge of the disc is located at $r_b = 5r_0$ in the equatorial plane, where $r_0$ is a reference length discussed below.
     The initial angular velocity of the disc is slightly sub-Keplerian, 
\begin{equation}
 \Omega\left|_{\theta=\pi/2} \right. = (1-0.003)\Omega_K(r) \hspace{1cm} r>r_b~,
\end{equation}
Inside of $r_b$, the matter rotates rigidly with angular velocity
\begin{equation}
 \Omega\left|_{\theta=\pi/2} \right. = (1-0.003)\Omega_K(r_b) \hspace{1cm} r  \leq r_b.
\end{equation}
The corotation radius $r_{\rm cr}$ is the radius where
the angular velocity of the disc equals that of the star; that is,
$r_{\rm cr}=(GM_*/\Omega_*^2)^{1/3}$.  
In this study we have chosen this radius to be the initial
inner radius of the disc with $r_{\rm cr} =5r_0$.

\subsection{Boundary Conditions}

Our simulation region has four boundaries:  The surface of the star, 
the midplane of the disc, the rotation axis, and the external boundary. 
For each dynamical variable we impose boundary conditions consistent with our physical
assumptions.

We assume axisymmetry as well as symmetry about the
equatorial plane. On the star and the external boundary we want to allow fluxes and so impose free boundary
conditions $\partial {\cal F}/\partial r=0$, where $\cal F$ are the dynamical variables. In 
addition, along the external boundary in the disc region $\theta =72^\circ-90^\circ$, we allow matter to inflow but the inflowing matter has zero magnetic flux. In the coronal region 
$\theta =0^\circ-72^\circ$ we allow matter, entropy and magnetic flux to exit the simulation region.

In addition to the boundary conditions, which are required for the well-posedness of our problem, we impose additional  conditions 
on variables to eliminate numerical artifacts in the simulations. For instance, we require that the radial velocity at the surface of the star be negative.   Thus there there is no outflow of matter, angular
momentum, or energy from the star.
We also require that along the external boundary the radial velocity be inwards inside the disc and
outwards outside the disc so the disc matter will tend to accrete and matter in the corona will tend to be ejected.

    Figure 2 shows the grid used in the described simulations.
There are $N_\theta =30$ constant width cells in the $\theta-$direction. 
In the $r-$direction, the $N_r =67$ grid cells increase in width 
as $dr_{j+1} =(1+0.0523)dr_j$ so as to give curvlinear rectangles
with approximately equal sides (Ustyugova et al. 2006).   The
dependence of our results on the grid resolution is discussed
in Appendix A.

\begin{figure}
                \centering
                \includegraphics[width=0.35\textwidth]{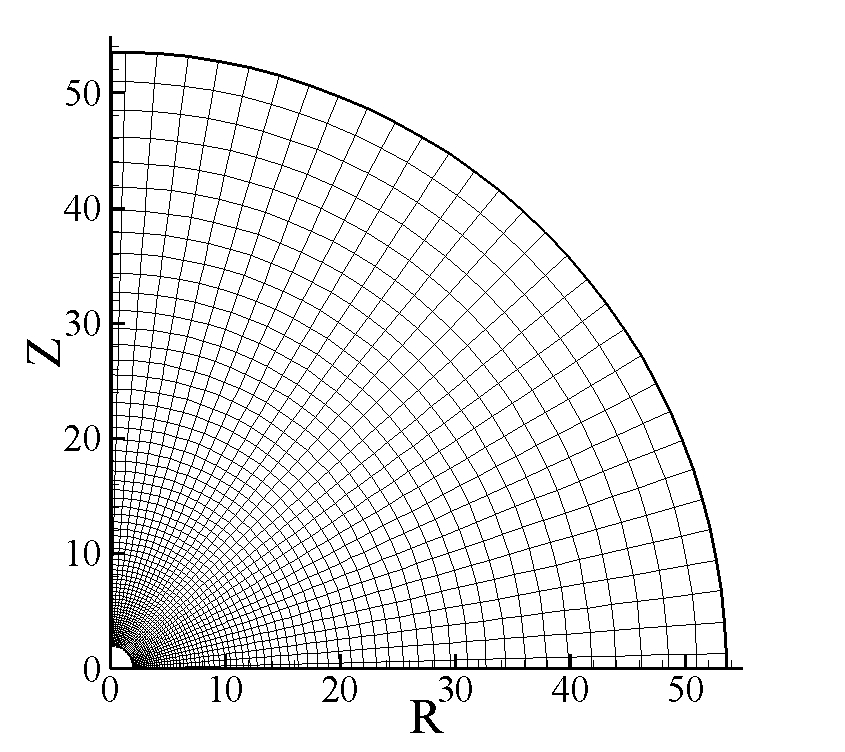}
        \caption{Grid used in the simulations.}
\label{fig:cartoon}
\end{figure}

\subsection{Dimensional Variables}

\begin{table}
\begin{center}
  \begin{tabular}{ | l  c  l |}
                                                                 \\\hline
Parameters           & Symbol    & Value                      \\ \hline \hline
   mass              & $M_*$     & $2.8\times10^{33}$ g       \\
   length            & $r_0$     & $1.0\times10^{6}$cm        \\
   magnetic field    & $B_0$     & $10^{8}$ G                 \\ \hline 
   time              & $P_0$       & $4.6\times10^{-4}$s       \\
   velocity          & $v_0$     & $1.4\times10^{10}$cm/s    \\ 
   density           & $\rho_0$   & $1.3\times 10^{-5} $ g/cm$^3$\\
   accretion rate & $\dot{M}_*$ & $2.8\times 10^{-9}$ $M_\odot$/yr \\
   disc power   &$\dot{E}_0$ & $1.66\times 10^{37}$ erg/s\\
        \hline \hline
  \end{tabular}
\end{center}
\caption{Mass, length, and magnetic field scales of interest and the corresponding scales of other derived quantities}
\label{table:units} 
\end{table}

   The MHD equations are written in dimensionless form 
so that the simulation results 
can be applied  to different types of stars. 
    The mass of the central star is taken as 
the reference unit of mass,  $M_0 = M_*$.
    The reference length, $r_0 $, is taken to
 be half the radius of the star.
    The initial inner radius of the disc is $r_b = 5 r_0$.
    The reference value for the
velocity is the Keplerian velocity at the radius $r_0$, 
$v_0 = (GM_0/r_0)^{1/2}$.  
    The dimensionless temperature is $T/v_0^2$.
     The
reference time-scale is the period of rotation at $r_0$, 
 $P_0 = 2\pi r_0/v_0$.
      From the MHD equations, we get the
relation $\rho_0 v_0^2 = B_0^2$,
where $B_0$  is a  reference magnetic field 
and $\rho_0$ is a reference  density both at
$r_0$. 
         We take the reference magnetic field $B_0$ to be such that
the reference density is appropriate for the considered star.
    The reference mass accretion rate is $\dot{M}_0 = \rho_0 v_0 r_0^2$.
The reference disc accretion power is $\dot{E}_0 = GM_0\dot{M}_0/r_0$.
    The initial dimensionless temperature in the disc is $T_{\rm{disc}} = (p/\rho)_{\rm{disc}} = 5\times10^{-4}$, and the initial temperature in the corona is $T_{\rm{cor}} = 
(p/\rho)_{\rm{cor}} = 0.5$.

Results obtained in dimensionless form can be applied to objects with 
widely different sizes and masses. 
    However, the present work focuses on neutron stars with the
typical values shown in Table 1.

\section{Results}

\begin{figure*}
                \centering
                \includegraphics[width=.92\textwidth]{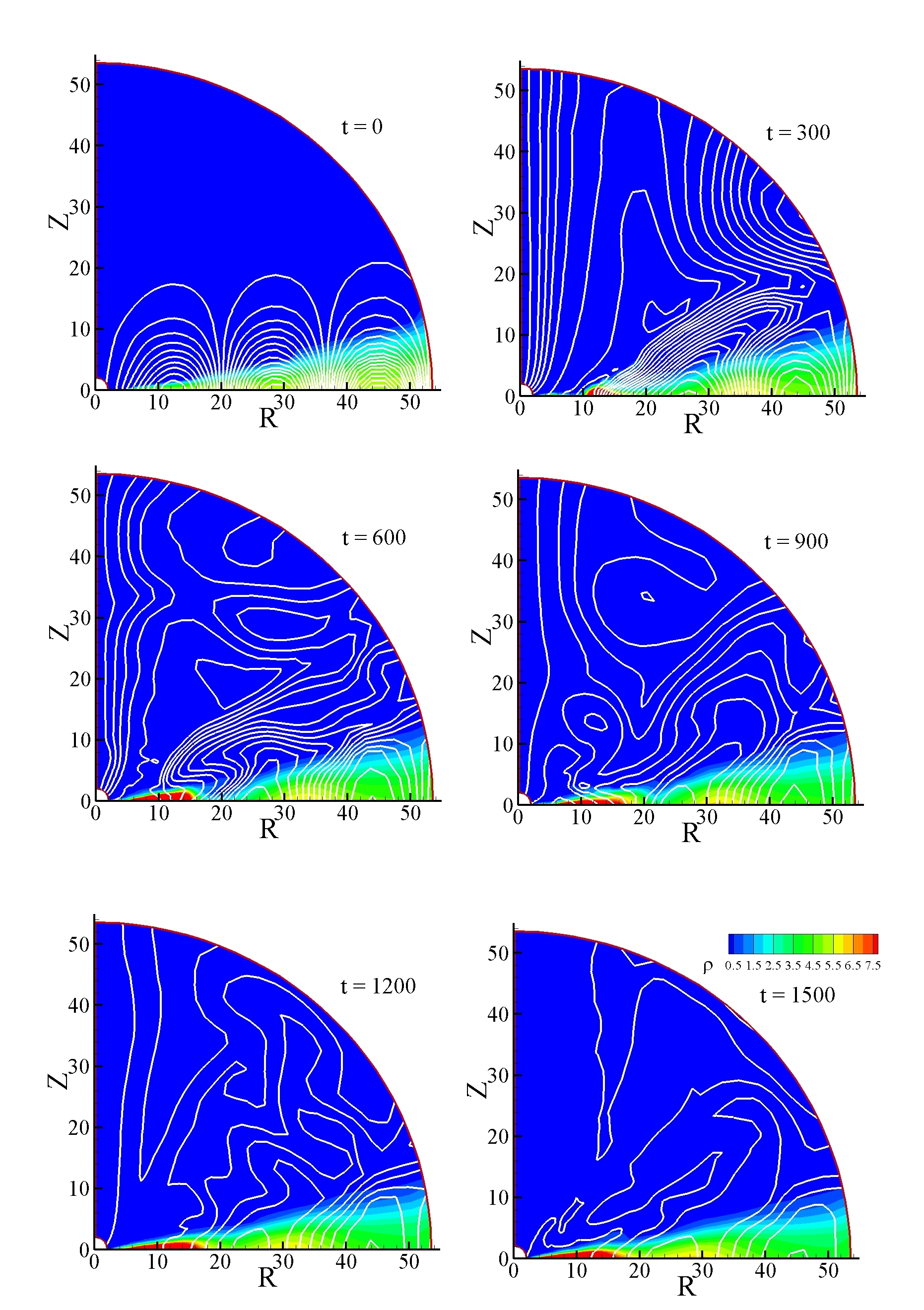}
        \caption{Plot of poloidal magnetic field lines (white) and matter density (colour) for $\alpha_{\nu}$ = 0.1 and $\alpha_{\eta}$ = 0.1 at  $t=0, ~=300,~ .., 1500$, where $t$ is the time measured in units of $P_0$ which is the period of the Keplerian orbit at the reference radius
$r_0$ (see \S 2.4).
By $t = 300$ sufficient matter has fallen into the star to drag in the magnetic field. The latter forms a well collimated jet along the axis and a wind along the  disc. This persists for some time, but eventually decays due to magnetic field reconnection  and annihilation.}
\label{fig:0}
\end{figure*}

        \begin{figure}
                \centering
                \includegraphics[width=.5\textwidth]{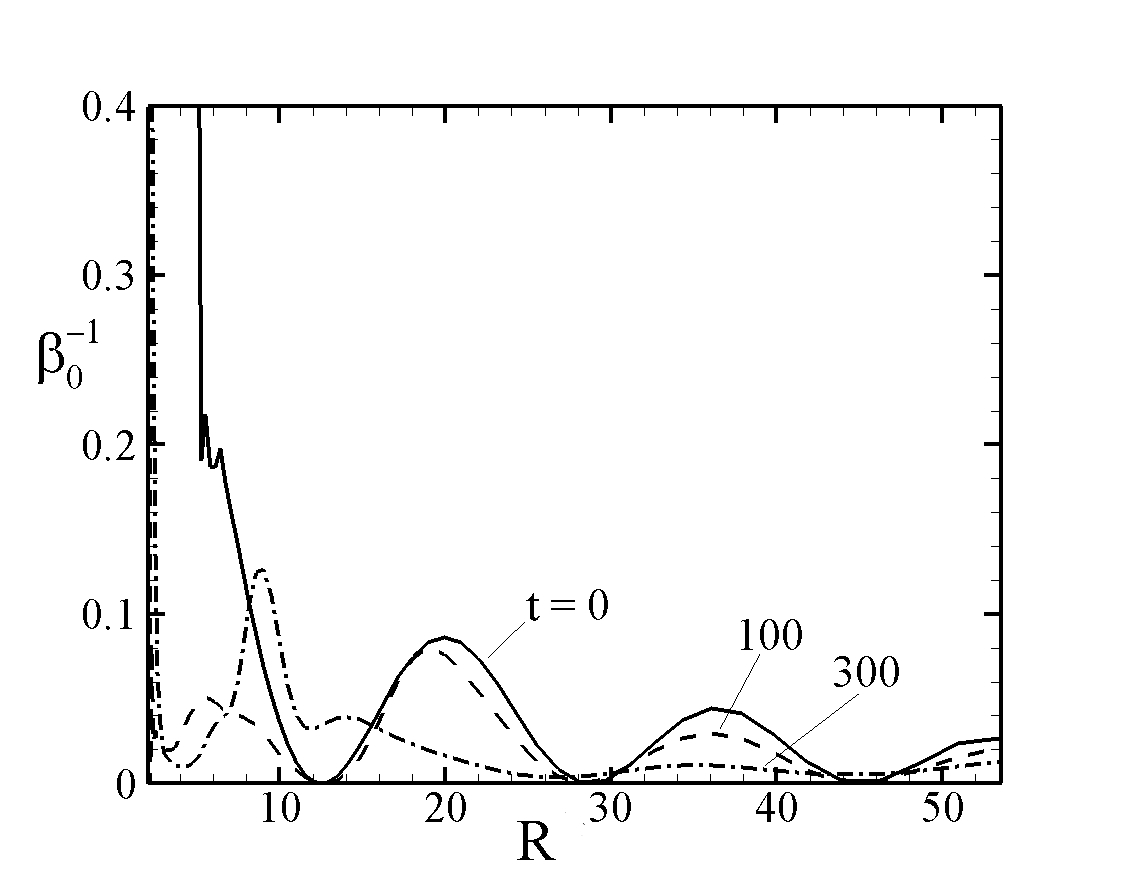}
                \label{fig:betaA}
        \caption{The ratio of magnetic pressure to the
plasma pressure  $\beta_0^{-1}$ in the disc midplane at $t=0$ (full), $ t = 100$ (dashed) and $t = 300$ (dot-dashed) for 
$\alpha_{\nu} = 0.1$ and $\alpha_{\eta}=0.1$. }
        \end{figure}

        \begin{figure}
                \centering
                \includegraphics[width=.5\textwidth]{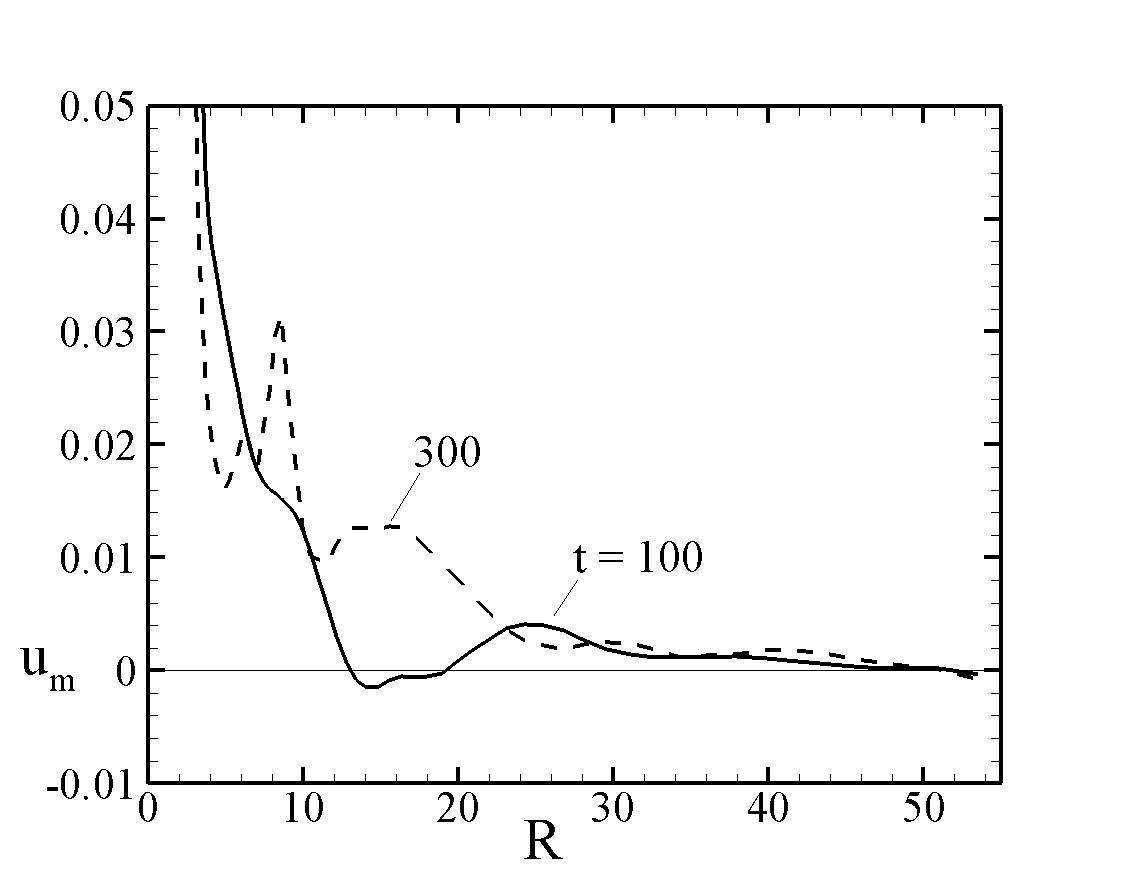}
                \label{fig:VR}
        \caption{Radial accretion speed of
the disc matter (equation 16) at $t = 100$ (full) 
and $t = 300$ (dashed) for $\alpha_{\nu} = 0.1$ and $\alpha_{\eta}=0.1$.}
        \end{figure}

        \begin{figure}
                \centering
                \includegraphics[width=.5\textwidth]{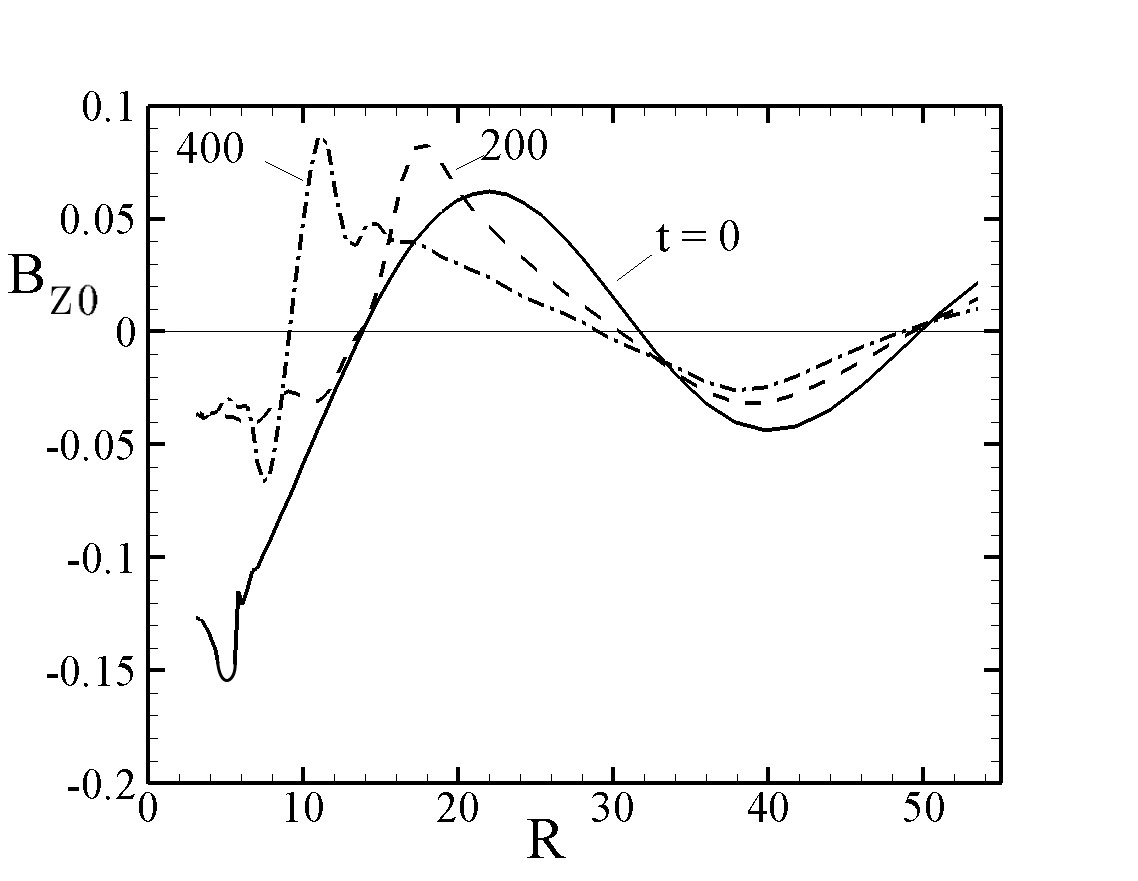}
                \label{fig:BZ}
        \caption{The magnetic field $B_Z(R,Z=0)$ along the disc
midplane at $t = 0$ (full), $t = 200$ (dashed) and $t = 400$ (dot-dashed) for $\alpha_{\nu} = 0.1$ and  $\alpha_{\eta}=0.1$.}
        \end{figure}

        \begin{figure}
                \centering
                \includegraphics[width=.5\textwidth]{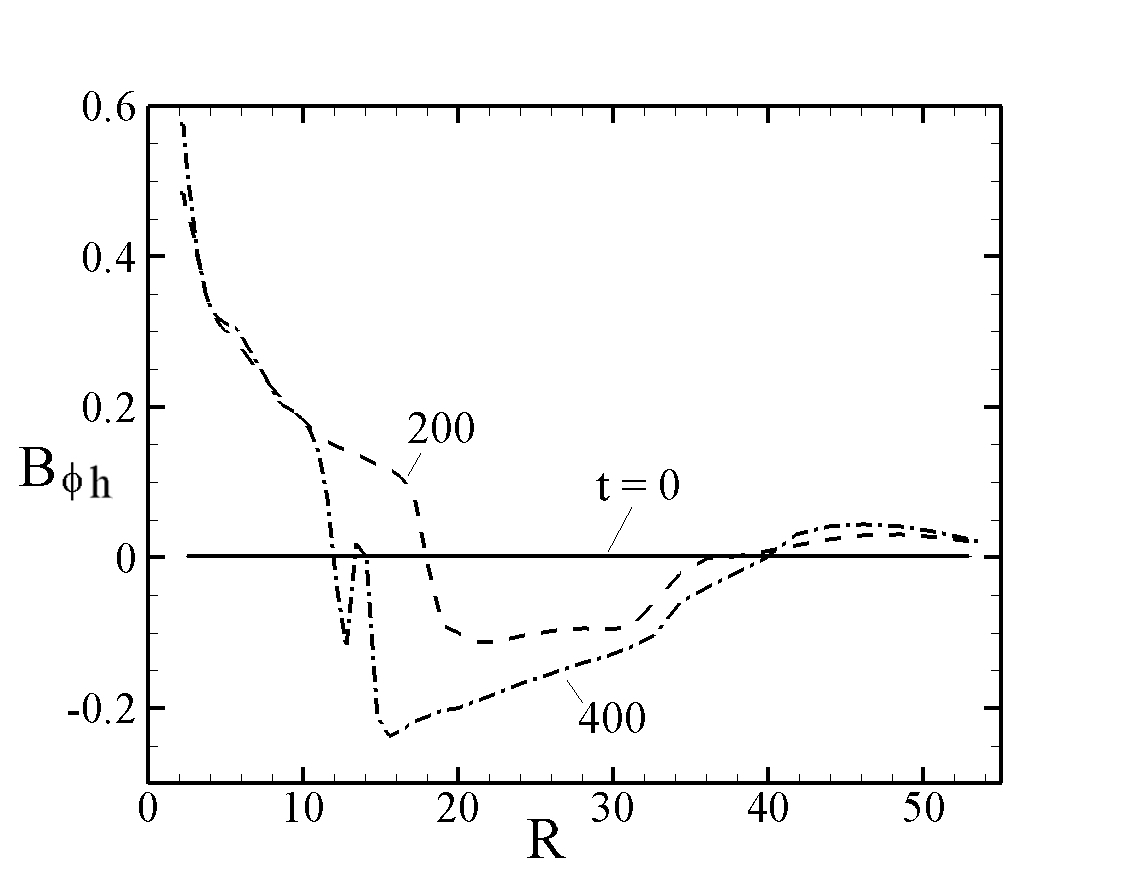}
                \label{fig:Bphi}
        \caption{The toroidal magnetic field $B_{\phi}(R,Z=h)$ 
at the disc surface  at $t = 0$ (full), $t = 200$ (dashed) and $t = 400$ (dot-dashed) for  $\alpha_{\nu} = 0.1$ and $\alpha_{\eta}=0.1$.}
        \end{figure}

        \begin{figure}
                \centering
                \includegraphics[width=.5\textwidth]{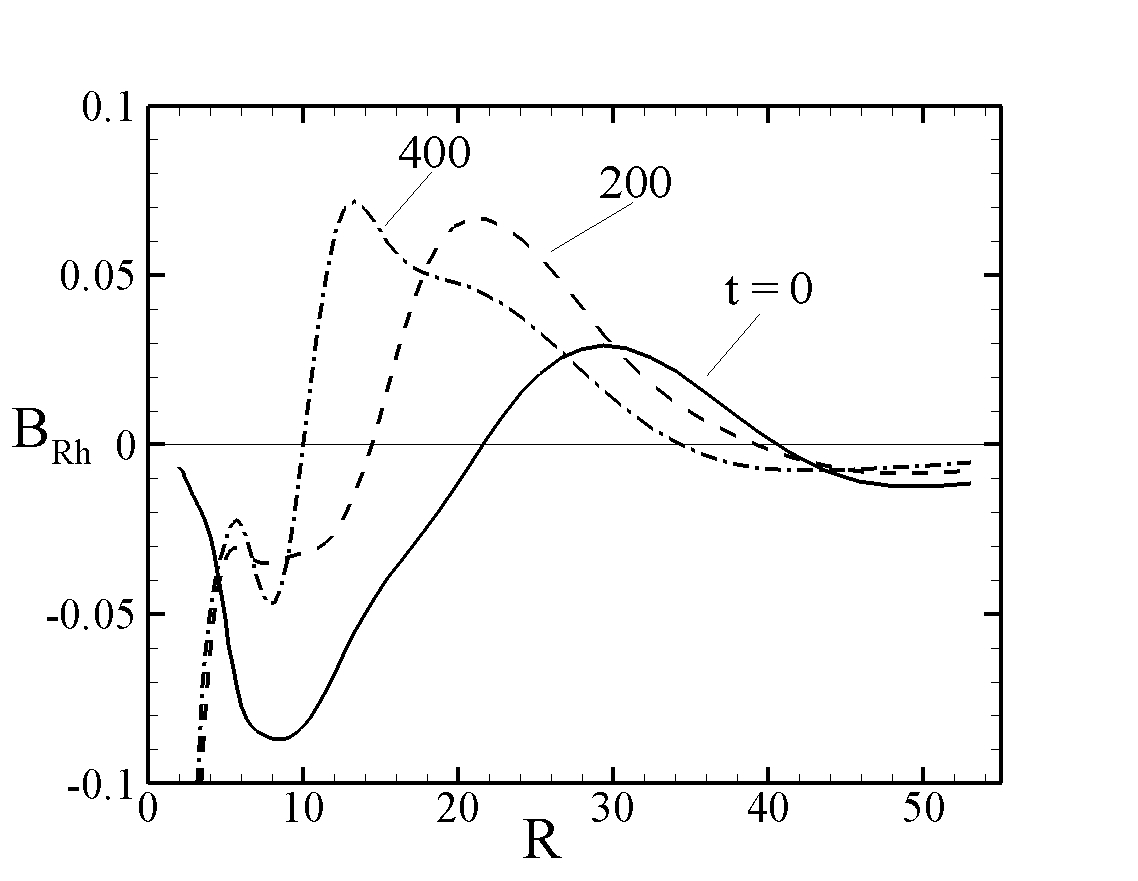}
                \label{fig:Bphi}
        \caption{The radial component of the magnetic field $B_{R}(R,Z=h)$ 
at the disc surface  at $t = 0$ (full), $t = 200$ (dashed) and $t = 400$ (dot-dashed) for  $\alpha_{\nu} = 0.1$ and $\alpha_{\eta}=0.1$.  This quantity has an important role in determining the diffusive advection of  the magnetic field $u_{B\eta}$ (equation 26).}
        \end{figure}

        \begin{figure}
                \centering
                \includegraphics[width=.5\textwidth]{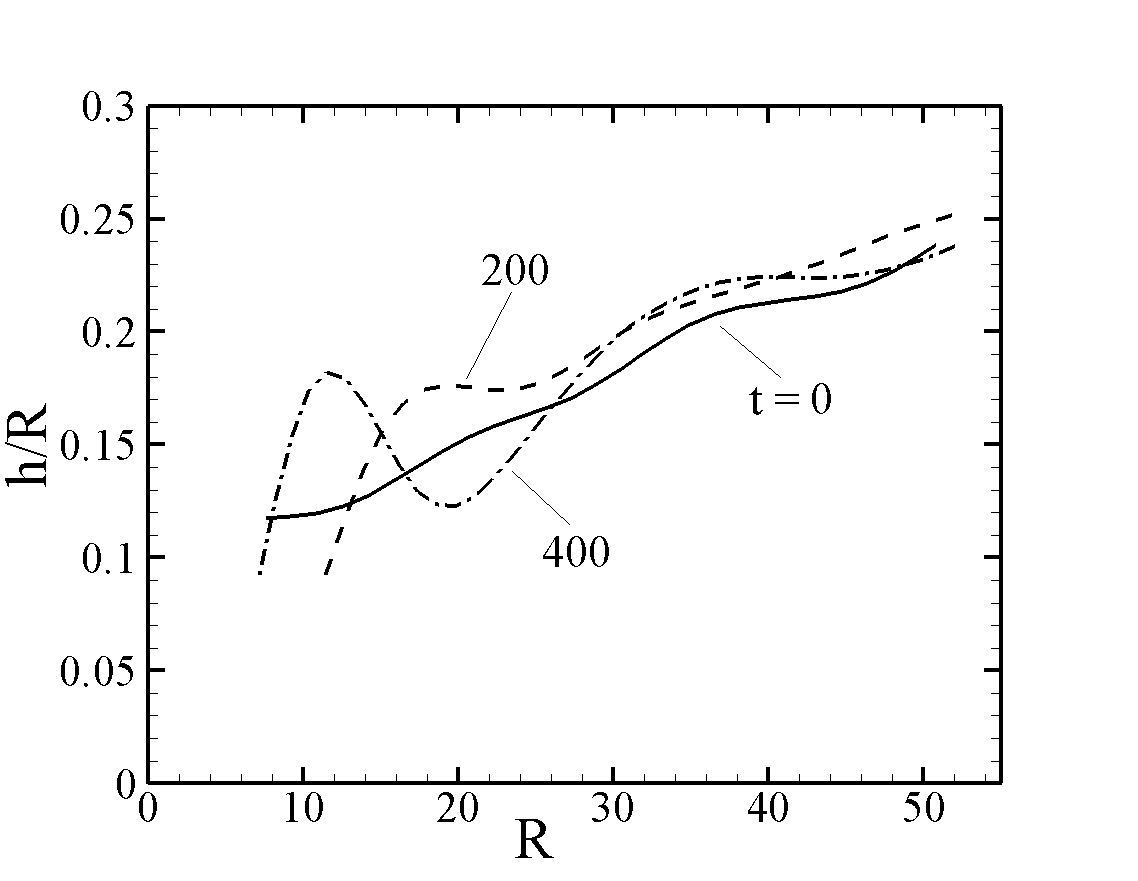}
                          \caption{The radio of the disc half thickness to the
                       radius $h/R$ as a function of $R$.}
        \end{figure}

\begin{figure}
\vspace{-0cm}
                \centering 
                \includegraphics[height=0.4\textwidth]{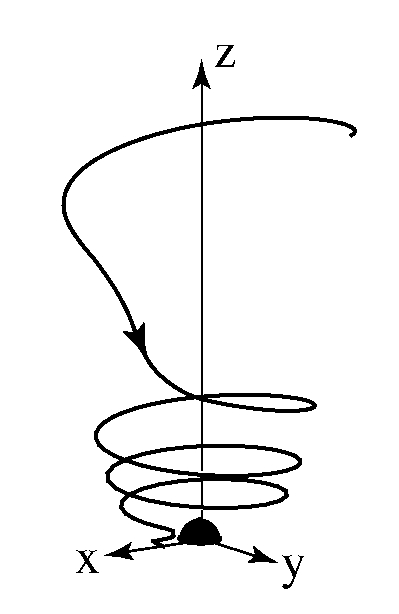} 
        \vspace{-0cm}
        \caption{Three dimensional view of a magnetic field
line originating from the disc at $R=6$ at time $t=300$.  The
twist of the field line is such that the field transports angular
momentum out of the disc.  That is, $B_Z B_\phi <0$.}
\label{fig:bfobzFig}
\end{figure}

We have carried out a large number of simulation runs for different values of the viscosity $0.05 \leq \alpha_{\nu} \leq 0.3$ and diffusivity $0.01 \leq \alpha_{\eta} \leq 0.3$ parameters and find that the simulations exhibit  similar qualitative behaviour.   
       Figure 3 shows the evolution of
the poloidal field projections for a representative case where $\alpha_\nu =0.1$
and $\alpha_\eta =0.1$.
    The field lines of the innermost loop
 are pulled in towards the star by the accreting
disc  matter. 
     When this loop reaches the star's surface it opens up.
     The inner half of the loop extends vertically upwards 
from the star,  supporting a magnetically dominated jet along
the $z-$axis.
     The outer half of the loop threading the disc projects outwards from
the disc at about $45^\circ$ to the disc normal,  
and it  supports a magnetic disc wind.    
    The middle and outer magnetic loops 
move inward only gradually, and they decrease in strength due
to field line  annihilation inside the disc.

Figure 4 shows the radial dependence  of 
the inverse ``plasma beta'' which is the
ratio of the magnetic pressure to the plasma pressure at the
disc midplane,  
\begin{equation}
\beta_0^{-1} ={ B_{Z0}^2 \over 8\pi p(R,0)}~,
\end{equation}
where $B_{Z0}\equiv B_Z(R,Z=0)$.
Note that $\beta_0^{-1} =(v_{A0}/c_{s0})^2/2$, where
$v_{A0}=B_{Z0}/\sqrt{4\pi \rho}$ is the midplane Alfv\'en
speed and $c_{s0}$ is the midplane sound speed.   
   For the assumed symmetry about the equatorial plane
$B_Z$ is the only non-vanishing field component at $Z=0$.
     Initially $\beta_0^{-1}$ is significantly less than unity over most of the disc ($R\gtrsim 5$).   Consequently, the magneto-rotational instability  would be expected
to occur in an actual disc with the same parameters (Balbus \& Hawley
1998).
     At later times, $[B_Z(R,0)]^2$ increases but the plasma pressure
 also increases so that the change in $\beta_0^{-1}$ are not
 very large.
 
 \subsection{Opening of Field Loops and Field Annihilation}
 
     The initial three  poloidal field loops ($t=0$ panel of Figure 3),
have footpoints at the approximate radii $R_k =5,~20, ~35, ~\&~ 50$
($k=1,..,4$).
     The time-dependent opening of large-scale magnetic field loops with
footpoints at different radii in a Keplerian disc has been well
analyzed (Newman, Newman, \& Lovelace 1992;  Lynden-Bell
\& Boily 1994; Romanova et al. 1998; Ustyugova et al. 2000;  Lovelace et al. 2002).
    We use the numerical condition of Lynden-Bell and Boily (1994)
 that a field loop opens after there is a differential rotation
 of the footpoints by $>3.63 $ radians.  
     For each loop, the differential
 rotation in a time $t_k$ is $\Delta \phi =t_k [\Omega_K(R_k)-
 \Omega_K(R_{k+1})]$.   With $\Delta\phi =3.63$
 we obtain the opening times for the three loops,
 $t_1 =40.8$, $t_2=329$, and $t_3= 760$ in our dimensionless
 units.  
      The observed rapid opening of the first loop (Figure 3) agrees
qualitatively with $t_1$.
    The fact the second loop does not open in a time $\sim t_2$ may be explained by the overlying magnetic field of the first loop.
     The long time $t_3$ required for the opening of the third loop
 means the diffusion of the magnetic field has sufficient time
 to cause significant  field annihilation.   
  
      The time scale for the field to diffuse over a distance
$\Delta R_k/2 = (R_{k+1}-R_k)/2$ can be estimated as
$\tau_k  \approx (\Delta R_k)^2/(4 \eta_t)$, where $\eta_t$ is evaluated
at $\overline{R}_k =(R_k+R_{k+1})/2$.
    For $\alpha_\eta =0.1$, we find $\tau_1\approx 2500$, $\tau_2\approx
 1700$,
 and $\tau_3\approx 1400$.  
       Note that $\tau_3$ for the outer  loop is less than the  duration
 of our runs so that the magnetic field  decays
 significantly before the end of the runs.


\begin{figure}
                \centering
                \includegraphics[width=0.5\textwidth]{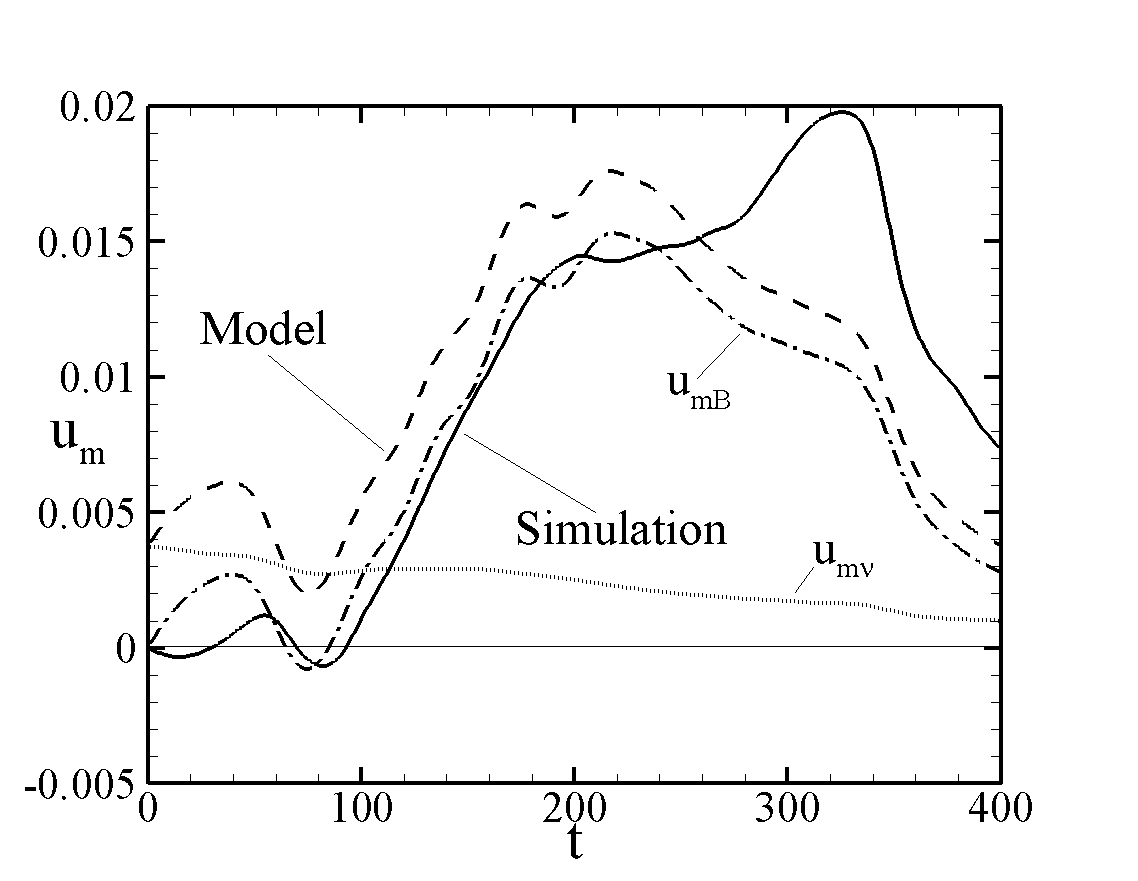}
        \caption{
The solid curve shows the average accretion speed of the disc matter at the maximum of $B_Z(R,Z=0)$ from the simulations for
$\alpha_\nu=0.1 =\alpha_\eta$.   The initial radius of the
maximum is  $R=22$.
 The dashed curve is from the  advection model (eqns. 19 and 20)) with parameters $C_1 = 0.05$ and $C_2=0.2$. 
     The model breaks down when the maximum approaches the star and interacts strongly with it.   This happens at different times for different viscosity and diffusivity values.}
\label{fig:model}
\end{figure}

\subsection{Matter Advection in the Disc}

    Figure 5 shows the average accretion speed $u_m$ of the disc matter
 as a function of $R$ at different times, where
\begin{equation}
u_m \equiv -{1\over \sigma(R)} \int_{-h}^h dZ \rho(R,Z) v_R(R,Z)~.
\end{equation} 
Here,
\begin{equation}
\sigma(R)= \int_{-h}^h dZ \rho(R,Z) ~,
\end{equation}
is the surface mass density of the disc.
    The conservation of mass gives 
   \begin{equation}
   {\partial (R\sigma)\over \partial t}
   -{\partial(R\sigma u_m)\over \partial R}=
    - {1\over \pi}{\delta \dot{M}_w \over \delta R}~,
    \end{equation}
where $\delta \dot{M}_w/\delta R = 2\pi R (\rho v_z)_{Z=h}$ is the
mass outflow rate from the top surface of the disc to the wind.
  The mass accretion rate to the star from the upper half-space
is $\dot{M}_* =(\pi R \sigma u_m)_{R=r_0}$.   
   We find that the ratio of the time-averaged 
 mass loss rate of the wind  is typically small
 compared to $\dot{M}_*$.

    Figure 6 shows the midplane magnetic field $B_Z(R,0)$   
 at different times.  
    Because of the assumed symmetry of the magnetic field about
 the equatorial plane,
 this is the only non-zero field component at $Z=0$.

   Matter advection in the disc is measured by
$u_m(R,t)$, which   is
determined by the disc's turbulent viscosity (which
causes the radial outflow of angular momentum) {\it and} by
the torque of the large-scale magnetic field (which causes
a vertical outflow of angular momentum to magnetic jets
or winds).
     This is described by a simple analytic model  where
\begin{equation}
{u}_{\rm{m}} = u_{m\nu} + u_{mB}~,
\end{equation}
where
\begin{equation}
u_{m\nu} =
3 C_1 \alpha_{\nu} \left( \frac{h}{R} \right)^2 v_K,~~
 u_{mB}={-C_2 B_{\phi h}B_{Z0}\over \pi v_K\sigma}~,
\end{equation}
(Lovelace et al. 2009; 1994),
where $h$ is the half-thickness of the disc, $v_K=(GM/R)^{1/2}$ is 
the  Keplerian velocity, $\sigma$ is the surface mass density of the disc,
 $B_{\phi h}=B_\phi(R,h)$ is the toroidal
magnetic field at the disc surface,
$B_{Z0} =B_Z(R,Z=0)$ is the midplane
magnetic field, and $C_1$ and $C_2$ are dimensionless constants
of the order of unity.   
    The first term  represents the accretion speed contribution due to
the turbulent viscosity of the disc, while the second term the accretion
speed contribution due to the outflow of angular momentum from
the disc surfaces due to the twisted magnetic field in the corona.

         Figure 7 shows the profiles of the
 toroidal magnetic field at the disc surface $B_\phi(R,h)$ at a
 sequence of times.   This quantity is important for
 the outflow of angular momentum from the disc which 
 in turn determines the magnetic contribution to the
 accretion speed $u_{mB}$ in equation (20).
 
     Figure 8 shows the radial profiles of $B_{Rh}$ at a sequence of times.
This quantity is important for the radial diffusion of the magnetic
field as discussed below in \S 3.3   Figure 9 shows the radial variation of
$h/R$ at a sequence of times. 
  
    Figure 10 shows that the twist of a sample  magnetic field line
above the disc is such that $B_{\phi h}B_{Z0} <0$ which
corresponds the to  outflow of angular momentum from the disc.

        \begin{figure}
                \centering
                \includegraphics[width=0.5\textwidth]{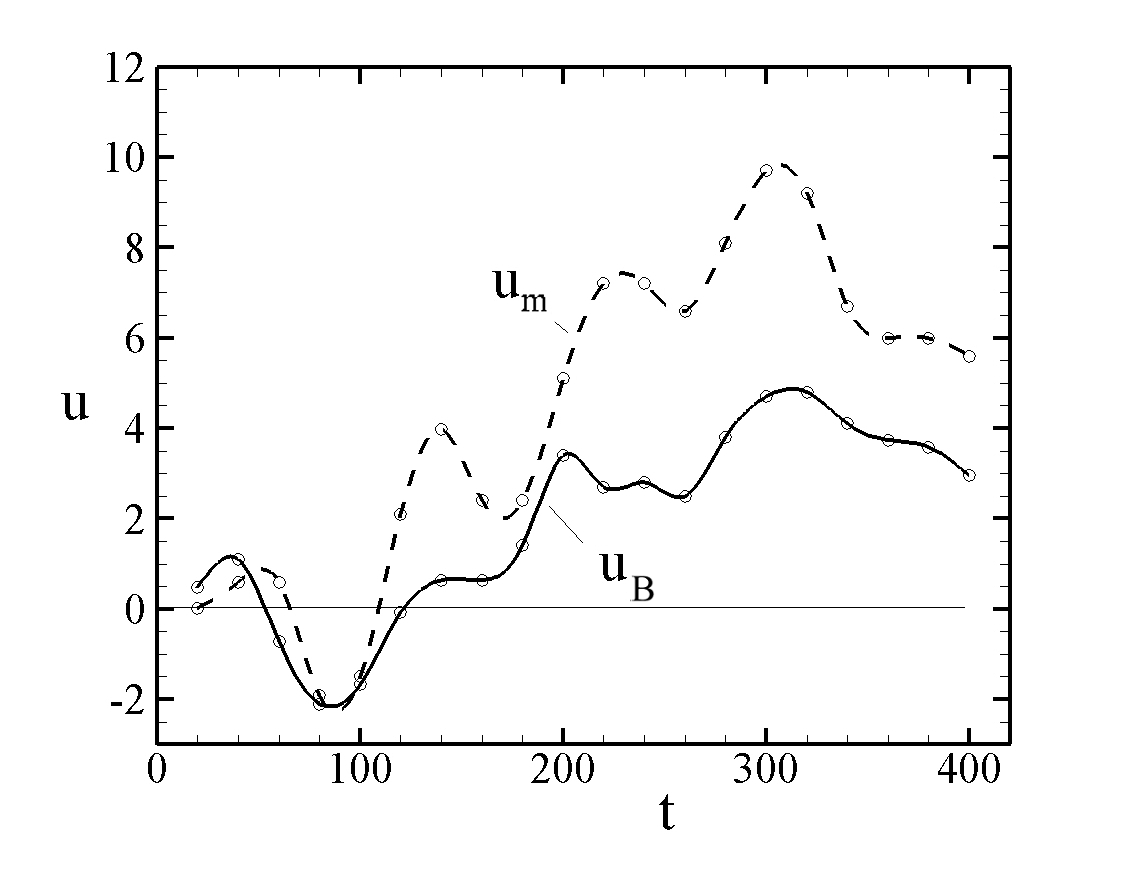}
       \caption{Accretion speed of the magnetic field $u_B$ 
in the equatorial plane derived
from the motion of the innermost  zero crossing of $B_Z(R,Z=0,t)$ and
the matter accretion speed $u_{\rm m}$ at the same location for
$\alpha_\nu=0.1=\alpha_\eta$.   The initial radius of zero
crossing is $R_{\rm O}(t=0)=13.7$.} 
        \end{figure}

We find that during early times ($t\lesssim 400$), the $B_Z(R,0)$   field 
 ``propagates'' inwards towards the star as
shown in Figure 6.
     We study this   motion by measuring the positions of 
the inner maximum of $B_Z(R,0)$ as a function of time. 
     The maximum moves inwards until interactions with the star 
cause a more complicated behavior.
     At the $B_Z(R,0)$ maximum at a given time $t$,
we calculate the accretion speed $u_{\rm m}(R,t)$ in our
simulation data using equation (16).   
      We also calculate $B_{\phi h}(R,t)$, $B_{Z0}(R,t)$, $h/R$, 
 and $\sigma(R,t)$ which permits us to
 compare the observed  accretion speed 
with  the prediction 
of the advection model (equations 19 and 20).
       We find that the reasonable values of
 $C_1 = 0.4$ and $C_2 = 0.2$ give satisfactory agreement between the model and our various simulation runs  for $t\lesssim 400$.
 
     Figure 11 shows the model and the measured simulation accretion speeds for sample cases.

\subsection{Magnetic Field Advection in the Disc}

     The advection of the poloidal magnetic field
is described by the equation
\begin{equation}
   {\partial (R\overline{B}_Z)\over \partial t}
   -{\partial(R\overline{B}_Z u_{Bi})\over \partial R}=
    {\partial \over \partial R}\left(-{\eta_t RB_{Rh} \over h}
    +\eta_t R {\partial \overline{B}_Z \over \partial R}\right)~,
\end{equation}
(Lovelace et al. 1994), where $B_{Rh} =B_R(R,Z=h)$. 
For simplicity we have neglected terms of
order $|\partial h/\partial R|$ relative to unity.   
     Here,
\begin{equation}
u_{Bi} \equiv - \int_{-h}^h dZ B_Z(R,Z) v_R(R,Z)
\bigg/\int_{-h}^h dZ B_Z(R,Z),
\end{equation}
is the magnetic field  advection speed of an ideal,  perfectly
 conducting disc ($\eta_t=0$).    
Note that the matter advection speed $u_m$ is a density weighted average over
the disc thickness
of $-v_R$ whereas $u_{Bi}$ is an average of $-v_R$ weighted by
$B_Z$.   For smooth profiles of $\rho$ and $B_Z$ the two speeds will be
be comparable.

     The vertical magnetic flux threading
 the disc inside the first O-point where $\overline{B}_Z(R_{\rm O},0)=0$  (or between successive O-points) decreases  in general with time due to the diffusivity.
From equation (21) we have
\begin{equation}
{d \over dt} \int_0^{R_{\rm O}} RdR \overline{B}_Z
=\left( -{\eta_t R B_{Rh} \over h} +
 \eta_t R{\partial \overline{B}_Z \over \partial R}\right)_{R=R_{\rm O}}~.
\end{equation}
  For example,  for $\overline{B}_Z < 0$ inside $R_{\rm O}$, both
terms on the right-hand side of equation (23) are seen to be positive
so that the magnitude of the flux decreases.

For $\eta_t>0$ the field advection speed is the sum of the ideal
and diffusive contributions,
\begin{equation}
u_B =u_{Bi} +u_{B\eta}~,
\end{equation}
with
\begin{equation}
   {\partial (R\overline{B}_Z)\over \partial t}
   -{\partial(R\overline{B}_Z u_{B})\over \partial R}=0~.
\end{equation}
Here
\begin{equation}
u_{B\eta}=-{\eta_t B_{Rh} \over h\overline{B}_Z}
    +{\eta_t \over \overline{B}_Z} {\partial \overline{B}_Z \over \partial R}
 \end{equation}
is the diffusive advection speed.

      From our simulation data we can calculate the advection speed of the magnetic field $u_{Bi}$ by tracking the location $R_{\rm O}(t)$  of the $B_Z(R,Z=0)$ zero crossings
 which occur at  ``O-points'' of the poloidal magnetic field ${\bf B}_p$.   
       Close to the
 zero crossing, $B_Z(R,0)= {\rm const} (R-R_{\rm O})$ is an odd function of $R-R_{\rm O}$.  Furthermore,
 $\overline{B}_Z$ is also an odd function of $R-R_O$.  
 Thus
 $u_{B\eta}$ is an odd function about the O-point proportional to
 $(R-R_{\rm O})^{-1}$ because of the $\overline{B}_Z$
 denominators in equation (26).   The magnetic field moves symmetrically 
 inward towards the O-point where it annihilates.
 The mathematical singularity is
 smoothed out by the finite grid.   The magnetic field inside a current-carrying 
resistive wire behaves in the same way.
 Consequently,  at $R=R_{\rm O}$
 we have $u_{B\eta} =0$.    The diffusivity has
 no influence on the motion of the O-point.  
 That is, $dR_{\rm O}/dt = -u_B = -u_{Bi}$.

     We can   compare field advection speed $u_B$ (at an O-point) to the matter 
advection  speed $u_{\rm m}$ at the same location.   
     As mentioned  these two speeds are expected to
 be comparable.
    
    Figure 12 shows sample comparisons of $u_B$ and $u_{\rm m}$.  
    For  a smaller diffusivity with
 fixed viscosity the difference between matter and field accretion
 speeds is smaller. 
     For a larger  viscosity relative to diffusivity, $u_m$ is significanty
 larger than $u_B$.

\subsection{Late times}

     At late times ($t \gtrsim1000$), the magnetic field decays appreciably
owing to reconnection and field annihilation.  
   Consequently the accretion speed is due mainly to the
disc viscosity,
\begin{equation}
{u}_{\rm m} \approx u_{m\nu}~.
\end{equation}
The mass accretion rate to the star  from the top half space is
$ \dot{M}_* =\pi (R\Sigma u_{\rm m})_*$, where $\Sigma$ is the surface
mass density of the disc and the asterisk subscript indicates
evaluation  outside the star.  
  
    Figure 13 shows the  time dependence of the accretion rate
to the star and the mass outflow rate in the wind for
two viscosity values and $\alpha_\eta =0.1$.     
    For $0.05 \leq \alpha_\nu \leq 0.3$ we find that $\dot{M}_*$ 
 at late times ($t>1000$) is approximately 
 proportional to $\alpha_\nu$.
   At late times $\dot{M}_*$ is independent of the
diffusivity $\alpha_\eta$.

\begin{figure*}
                \centering
                \includegraphics[width=\textwidth]{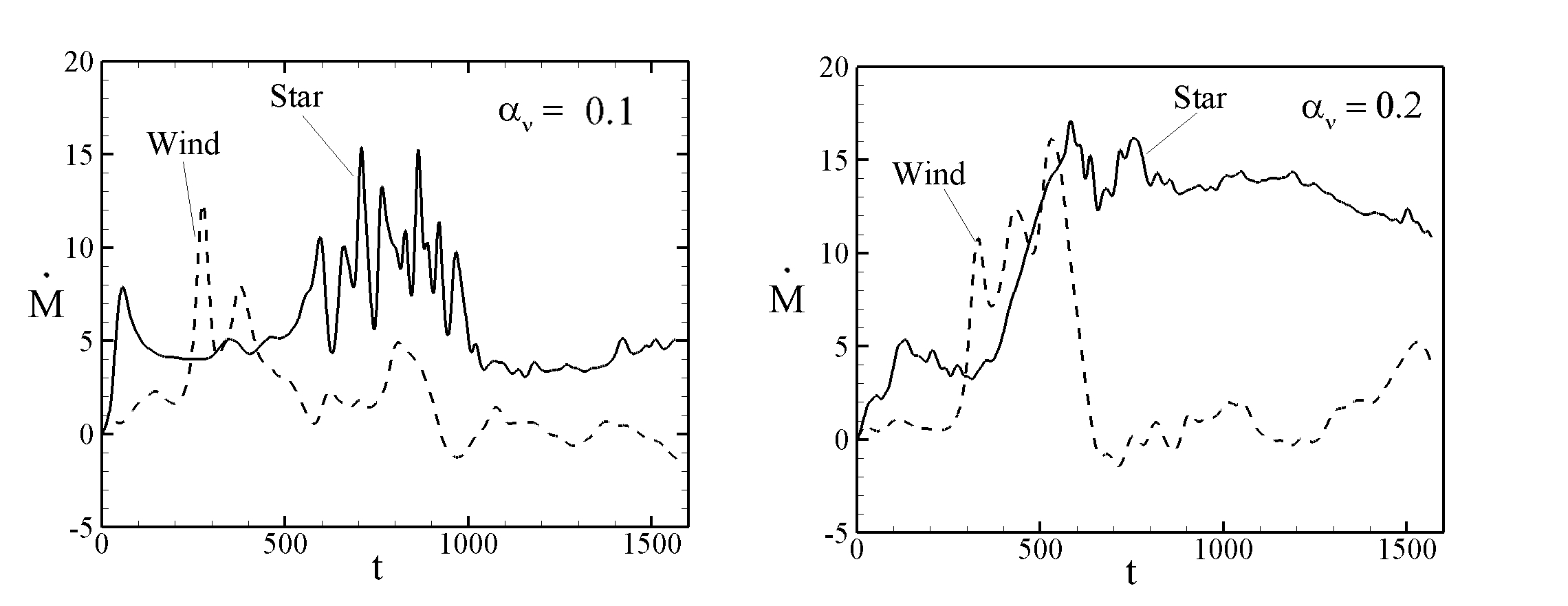}
        \caption{Dimensionless mass accretion rate to the star $\dot{M}_*$
and mass loss rate to the wind $\dot{M}_w$      for two viscosities with same diffusivity $\alpha_\eta =0.1$.} 
\label{fig:MdotALPHA}
\end{figure*}

\subsection{Jet and Wind}

    In this work we observe both a collimated jet along the
$Z-$axis  and an uncollimated disc wind.

\subsubsection{Jet}
The fluxes of angular momentum and energy
through the spherical surface $[r=44,~ 0^{\circ} \leq \theta \leq 21^{\circ}]$ are shown in Figure 14
     The angular momentum flux can
be separated into a part from the matter and a part due to the magnetic field,
\begin{equation}
\dot{L} = \dot{L}_m + \dot{L}_f = \int d{\bf S} \ \cdot \left(\rho r\sin(\theta)v_{\phi} \mathbf{v}_{p}  -  \frac{r\sin(\theta)B_{\phi}\mathbf{B}_p}{4\pi} \right) . 
\end{equation}
Similarly, the energy flux can be separated into  contributions carried
by the  matter and that carried by
the Poynting flux,
\begin{equation}
\dot{E} = \dot{E}_m + \dot{E}_f = \int d{\bf S} \ \cdot \left( \frac{1}{2} \rho {\bf v}^2 \mathbf{v}_p + \frac{c}{4\pi}\mathbf{E} \times \mathbf{B}\right) . 
\end{equation}
The jet is strongly dominated by the electromagnetic
field:  The angular momentum flux is carried predominantly by the magnetic
field and the energy flux is carried predominantly by the Poynting flux.
    Such jets were hypothesized by Lovelace (1976) and first observed in
axisymmetric MHD simulations by Ustyugova et al. (2000).

\begin{figure*}
                \centering
                \includegraphics[width=\textwidth]{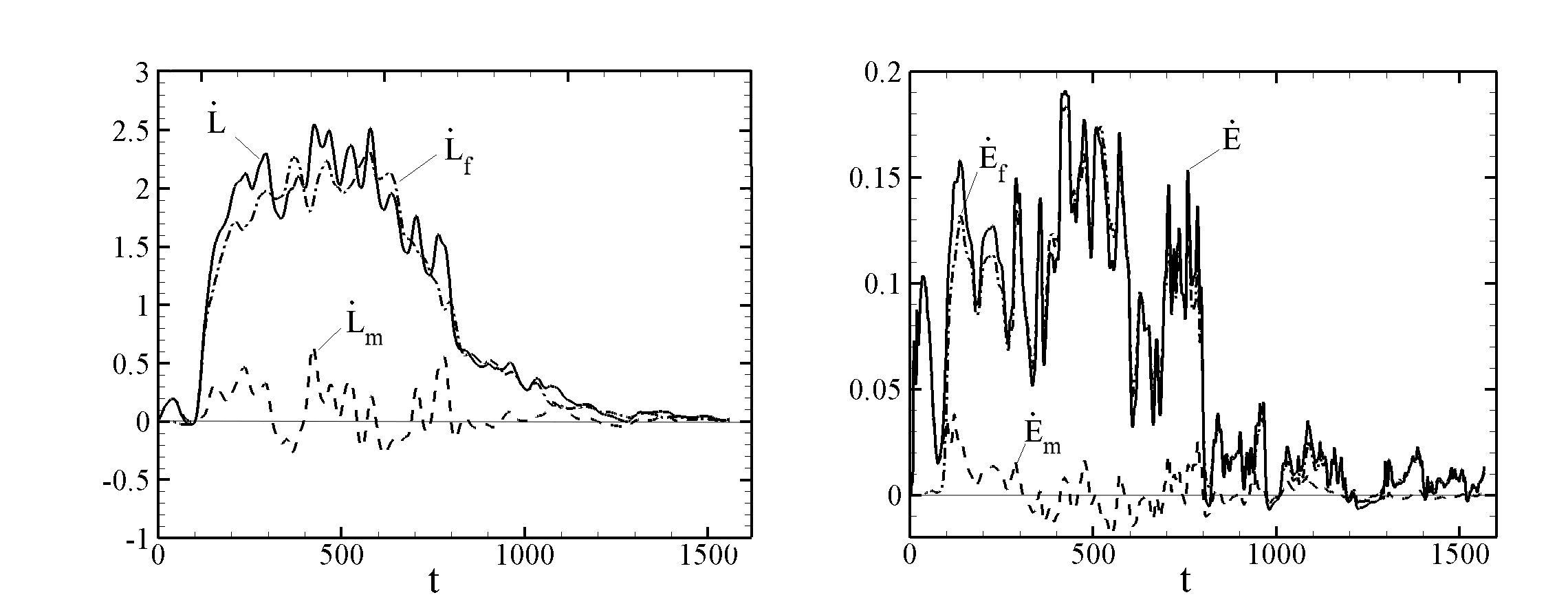}
        \caption{- \emph{Left panel}: The jet angular momentum flux carried
by the matter (dashed curve), the magnetic field (dot-dashed
curve), and the total flux (solid curve) for $\alpha_\nu=0.1$ and $\alpha_\eta=0.1$.
    \emph{Right-panel}: The jet energy flux carried by the
matter (dashed curve), 
the Poynting flux (dot-dashed curve), and the total flux (solid curve)
for $\alpha_\nu=0.1$ and  $\alpha_\eta=0.1$.
   The jet is strongly dominated by the Poynting flux.}
\label{fig:jetbreakdown}
\end{figure*} 

\subsubsection{Disc Wind}

The rates of energy, angular momentum and mass flux through the surface $[r=44, ~21^{\circ} \leq \theta \leq 72^{\circ}]$ is shown in Figure 15.
We have chosen 
the upper bound for $\theta$ by requiring that the wind stay outside the disc. 

     Figure 14 shows the different components of the wind angular momentum flux and the components of the energy flux.  The angular momentum and energy fluxes are dominated by the matter components.   This is the opposite
of the case for the jet.

   Figure 14 shows the jet and wind total energy fluxes
normalized to the ``accretion power'' 
$\dot{E}_{\rm acc} = GM_*\dot{M}_*/(2r_*)$.   
The large initial values of the ratios results from the fact
that $\dot{M}_*$ is initially zero.   
    Note that the two ratios are comparable.

\begin{figure*}
                \centering
                \includegraphics[width=\textwidth]{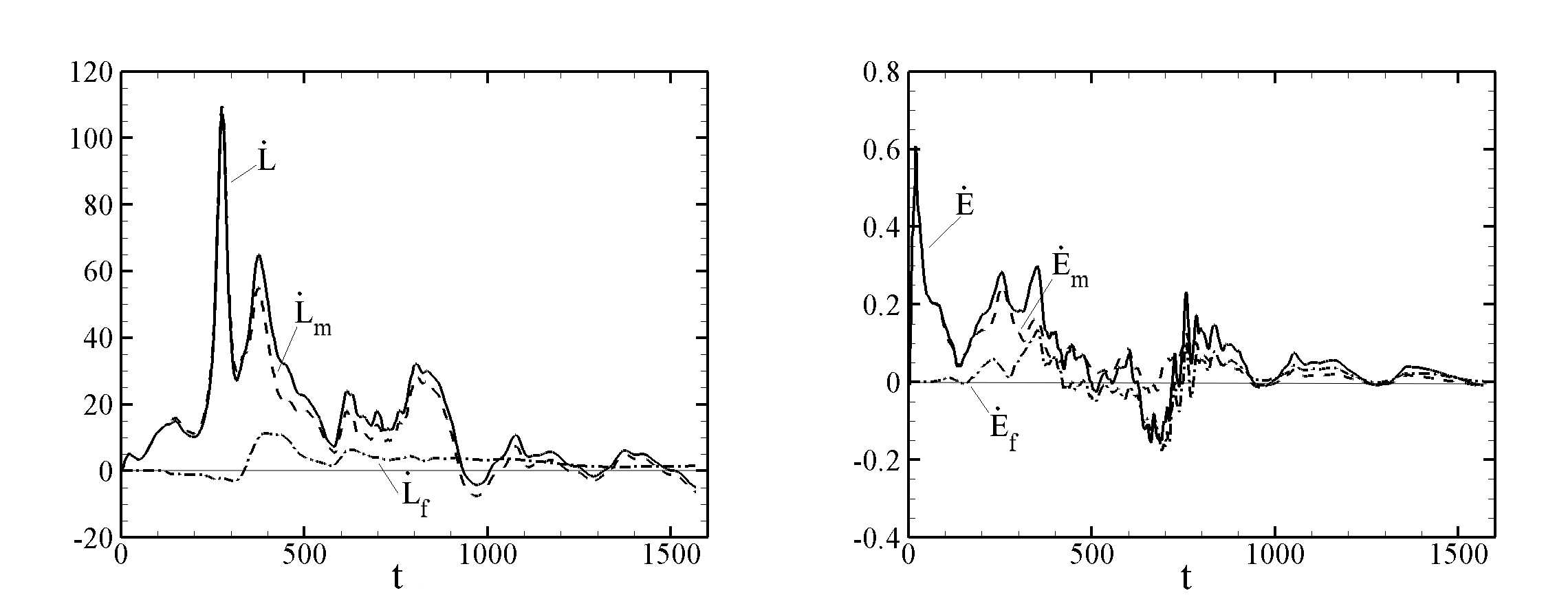}
       \caption{- \emph{Left panel}: The wind angular momentum flux carried by the matter (dashed curve), the magnetic field (dot-dashed
 curve),  and the total  (solid curve) for $\alpha_\nu=0.1$ and $\alpha_\eta=0.1$.
  \emph{Right panel}:   The wind energy flux carried by the
  matter (dashed curve), 
the Poynting flux (dot-dashed curve), and the 
total flux(solid curve) for $\alpha_\nu=0.1$ and $\alpha_\eta=0.1$.
    The wind is strongly matter dominated.}
\label{fig:windflux}
\end{figure*} 

\begin{figure*}
                \centering
                \includegraphics[width=\textwidth]{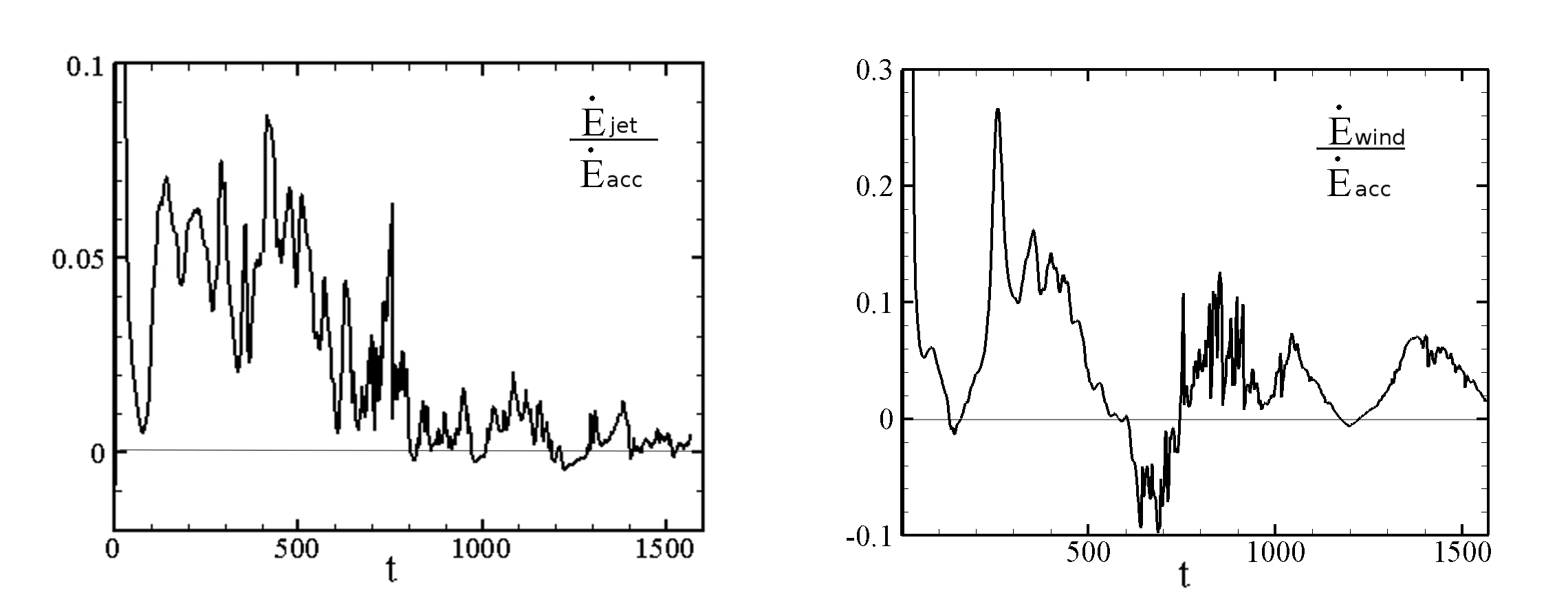}
       \caption{- \emph{Left panel}:   The total jet energy flux normalized
 by the  ``accretion power''  $\dot{E}_{\rm acc}=GM_*\dot{M}_*/(2r_*)$
 for $\alpha_\nu=0.1$ and $\alpha_\eta=0.1$.
         \emph{Right panel}:   The total wind power also normalized by the
accretion power $\dot{E}_{\rm acc}$ for $\alpha_\nu=0.1$ and $\alpha_\eta=0.1$.
The large initial values of the ratios is due to the fact that the 
$\dot{M}_*$ 
is initially zero.   The fact that $\dot{E}_{\rm wind}$ is temporarily negative
is due to the fixed partition of the range of $\theta$ into wind and disc
regions. }
\label{fig:windflux}
\end{figure*} 

\section{Conclusions}

   We have analyzed a set of axisymmetric MHD simulations
of the evolution of a turbulent/diffusive accretion disc 
initially threaded by a weak  magnetic field with midplane plasma
beta $\beta_0$ is significantly larger than unity.  
The viscosity and magnetic diffusivity are modeled by
two $\alpha$ parameters, one for the viscosity $\alpha_\nu$
and the other for the diffusivity $\alpha_\eta$.  
The coronal region above 
the  disc is treated using ideal MHD.   
      The  initial  magnetic field is taken to consist of three poloidal field loops threading the disc between its initial inner radius and to its ten times larger outer radius. 
 This field configuration allows the derivation of the 
advection speed of the magnetic field.

     Recent  theoretical studies discussed the
importance of the magnetic field extending from a turbulent
disc into a low density non-turbulent/highly conducting
corona (Bisnovatyi-Kogan \& Lovelace 2007;
Rothstein \& Lovelace 2008; Lovelace et al. 2009;
Bisnovatyi-Kogan \& Lovelace 2012;  Guilet \& Ogilvie 2012, 2013).   
     These treatments  all considered stationary or
quasi-stationary conditions {\it and} a disc threaded by a 
poloidal magnetic field of a single polarity. 
     In contrast the simulations discussed here are strongly time-dependent
and involve multiple poloidal field polarities in different regions
of the disc.      Consequently, a direct comparison of the theory
and simulations is not possible.
       The simulations clearly show the inward advection of the magnetic field at about the same
 speed as the matter advection  before the field decays by annihilation.

    At early times ($t \lesssim 400$), 
we find that the innermost
field loop twists and its field lines become open. 
    For the different field loops we estimate  two important
time scales:  One is the time scale for each loop to open due
to differential rotation of its foot points, and the other
is the field annihilation time scale  owing to the disc's magnetic
diffusivity.   The innermost field loop opens rapidly  before
there is significant annihilation.    On the other hand the outer
loop decays significantly before there is time for it to open.
    The twisting of
the opened field lines of the inner loop leads to the formation of {\it both} an inner collimated magnetically dominated jet and at larger distances from the axis a matter dominated uncollimated wind.   
    For later times ($>1000$), the strength of the
magnetic field decreases owing to field reconnection and
annihilation in the disc.
   For the early times, we have derived from the simulations both the matter accretion speed in the disc $u_{\rm m}$ and the accretion speed of the magnetic field $u_B$.   
     We show that the derived  $u_{\rm m}$  agrees approximately with the predictions of a model where the accretion speed is the sum of a contribution due the disc's viscosity (which gives a radial outflow of angular momentum in the disc) and a term due to the twisted magnetic field at the disc's surface (which gives
a vertical outflow of angular momentum) (Lovelace et al. 2009; 1994).
   At later times the magnetic contribution to $u_{\rm m}$ becomes
small compared with the viscous contribution.
   Also for early times we find that $u_{\rm m}$ is larger than the magnetic field accretion speed $u_B$ by a factor $\sim 2$ for
 the case where $\alpha_\nu =0.1=\alpha_\eta$.

\section*{Acknowledgments}

    We thank an anonymous  referee for valuable criticism which helped to improve this work.
This research was supported in part by NSF grant AST-1008636
and by a NASA ATP grant NNX10AF63G.


\appendix
\section{Dependence on Grid resolution}

    We have tested the dependence of our  results on the
grid by running higher resolution cases compared
with resolution used for this study, ($N_\theta,~N_R )=(31,~67)$.
  We have run cases with $(41,~87)$ and
$(50,~100)$.
Figure A1 shows the radial dependence of the mid
plane magnetic field $B_Z(R,Z=0)$ at $t=300$
for the three grid resolutions.  
     The radius of the first zero crossing of $B_Z(R,0)$ decreases
by about $14\%$ going from the low to the intermediate resolution.
  It decreases   by a further $2\%$ going from the intermediate to
the high resolution grid.  Thus the convergence is rapid.  
      This indicates the accuracy of the field
advection speed $u_B$ in Figure 12 and suggests that the actual
speed is higher by about $16\%$.

\begin{figure}
                \centering
                \includegraphics[width=0.45\textwidth]{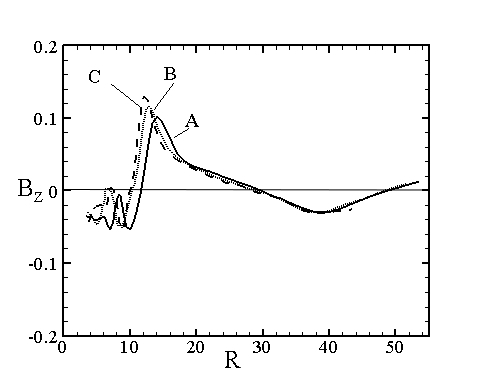}
        \caption{Radial dependences of $B_Z(R,Z=0)$ for 
grid  resolutions $(N_\theta,N_R)$ of $A=(31,67)$, $B=(41,87)$,
and $C=(50,100)$ at $t=300$
for $\alpha_\nu=0.1$ and $\alpha_\eta=0.1$.
 }
\label{fig:cartoon}
\end{figure}

\label{lastpage}

\end{document}